\documentclass[aps,prd]{revtex4}
\usepackage{eurosym}
\usepackage{amssymb}
\usepackage{amsfonts}
\usepackage{amsmath}
\usepackage{graphicx}
\usepackage{graphics}
\usepackage{eurosym}
\usepackage{xcolor}
\usepackage{amsmath,array}
\usepackage[utf8x]{inputenc}
\usepackage{changes}

\begin{document}

\title{Linear Stability of Mandal-Sengupta-Wadia Black Holes}
\author{H. G\"{u}rsel}
\email{huriye.gursel@emu.edu.tr}
\author{G. Tokg\"{o}z}
\email{gulnihal.tokgoz@emu.edu.tr}
\author{I. Sakalli}
\email{izzet.sakalli@emu.edu.tr}
\affiliation{Physics Department, Faculty of Arts and Sciences, Eastern Mediterranean University, Famagusta, Northern
Cyprus, Mersin 10, Turkey}
\date{\today }

\begin{abstract}
In this letter, the linear stability of static Mandal-Sengupta-Wadia (MSW) black holes in $(2+1)$-dimensional gravity against circularly symmetric perturbations is studied. Our analysis only applies to non-extremal configurations, thus it leaves
out the case of the extremal (2+1) MSW solution. The associated fields are assumed to have small
perturbations in these static backgrounds. We then consider the dilaton
equation and specific components of the linearized Einstein equations. The
resulting effective Klein-Gordon equation is reduced to the Schr\"{o}dinger like wave equation with the
associated effective potential. Finally, it is shown that MSW
black holes are stable against to the small time-dependent perturbations.
\end{abstract}

\keywords{Linear Stability, Dilaton, MSW Black Hole, Perturbation,
Klein-Gordon equation}
\pacs{04.20.Gz, 04.25.Nx, 04.40.Nr}
\maketitle

\section{Introduction}

One of the compelling problems in black hole physics is stability (for a
review see \cite{Dafermos} and references therein). The stability of a black
hole presents an ideal theoretical test-bed for any gravity theory \cite%
{Berti}. In the pioneering works on black hole stability \cite%
{STABL1,STABL2,STABL3,STABL4}, the linear stability of gravitational
perturbations of the Schwarzschild black hole was comprehensively studied.
The mode-stability of the Kerr black hole was first proved
in \cite{KERRstb}, which employed the Teukolsky equation \cite{Chandram}.
Recent developments have shown that there is a conserved and positive
definite energy in the axially symmetric linear gravitational perturbations
of the extreme Kerr black hole \cite{extreme}. This work stipulates the
framework of linear stability in axial symmetry. 
Although there exist worthwhile studies in literature regarding the non-linear
stability of black holes \cite{Hintz1,Hintz2}, further progress in the
concerned topic is required. In fact, linear stability studies are expected
to clarify the non-linear stability problem. However, this
is conditional on techniques that can be suitably extended to the non-linear
regime \cite{instability}. In general, no-hair theorems (NHTs) \cite{NHT}
are about the existence of black hole solutions and they do not consider the
stability of black holes. However, we know from the literature that some
black hole solutions have been rejected due to their instability under
perturbation \cite{Kanti}. As a matter of fact, there is a sort of consensus
among physicists that a new NHT is more likely to be accepted when the
stability of a black hole solution is assured. On the other hand, many black
holes that are well known in the literature and proved by NHTs have not been
tested for stability. For example, from the literature, the linear stability
of the famous three dimensional MSW black hole solution \cite{MSW}, which is
a solution in three dimensions from the classical dilaton system of Chan and
Mann \cite{ChanMann}, has not been studied before. Like the Ba$\widetilde{n}$%
ados-Teitelboim-Zanelli (BTZ) black hole model \cite{BTZbh}, which is a well-known toy model of three-dimensional black holes using general relativity
theory, MSW black holes have attracted much attention. For instance, the
problems of spectroscopy, thermodynamics, Hawking radiation, quasinormal
modes, and scalar perturbations for this black hole have been extensively
studied in \cite{msw1,msw2,msw3,msw4,msw5}. It is worth noting that the
studies on the $(2+1)$-dimensional black holes have the potential to
generate valuable insight into many conceptual issues that arise in $(3+1)$%
-dimensional black holes. Among those studies, Cardoso and Lemos \cite%
{CLBTZ} showed that quasinormal modes of the BTZ black hole for which the
frequencies are no longer pure real which means that the system is
losing energy. Therefore, the QNMs dominate the signal during the intermediate stages
of the perturbation, being therefore extremely important. In line with the
study \cite{Clem}, the main aim of the present letter is to address the
stability problem of MSW black holes, as hinted above. To this end, MSW
dilaton black holes are examined to see if they are stable against small
spacetime-dependent perturbations. The letter is laid out as follows. In
Sect. II, we review MSW black holes and their characteristic features. The
stability of the charged MSW static solutions against small perturbations is
studied in Sect. III. Finally, in Sect. IV we draw conclusions. (Throughout
the letter, natural units with $c=G=k_{B}=\hbar =1$ are used).

\section{MSW Black Holes}

MSW black holes redound to investigations in which the concepts of general
relativity and string theory can be linked. Examining the actions of
Einstein-Maxwell-dilaton and string theories \cite{Horow}, and using
conformal field transformations to relate them to each other, the
unification of these two theories can be achieved \cite{ChanMann}.
Throughout this section, we focus on MSW black hole solutions from the
general relativity perspective only.

$(2+1)$-dimensional Einstein-Maxwell-dilaton action is given by \cite%
{ChanMann}

\begin{equation}
S_{EMD}=\int d^{3}x\sqrt{-g}\left[ R-\frac{B}{2}\left( \nabla \phi \right)
^{2}-\exp \left( -4a\phi \right) F_{\mu \nu }F^{\mu \nu }+2\exp \left( b\phi
\right) \Lambda \right]  \label{1}
\end{equation}%}
in which $\phi $ and $F_{\mu \nu }$ represent the fields of concern (dilaton
and Maxwell fields, respectively), $a$, $b,$ and $B$ are arbitrary
constants, and $\Lambda $ stands for the cosmological constant. If we
perform variations in the metric, gauge, and dilaton field, we obtain the
following equations of motion:

\begin{equation}
R_{\mu \nu }=\frac{B}{2}\nabla _{\mu }\phi \nabla _{\nu }\phi +\exp \left(
-4a\phi \right) \left( -g_{\mu \nu }F^{2}+2F_{\mu }^{\alpha }F_{\nu \alpha
}\right) -2g_{\mu \nu }\exp \left( b\phi \right) \Lambda ,  \label{2}
\end{equation}

\begin{equation}
\nabla ^{\mu }\left( \exp \left( -4a\phi \right) F_{\mu \nu }\right) =0,
\label{3}
\end{equation}

\begin{equation}
\frac{B}{2}\left( \nabla ^{\mu }\nabla _{\mu }\phi \right) +2a\exp \left(
-4a\phi \right) F^{2}+b\exp \left( b\phi \right) \Lambda =0.  \label{4}
\end{equation}%
The above field equations with $b=4a=\frac{B}{2}=4$, and $\phi =-\frac{1}{4}%
\ln \left( \frac{r}{\beta }\right) $ \{$\beta :$constant\} lead to the
charged MSW black hole solutions, which can be stated as \cite{ChanMann}

\begin{equation}
ds^{2}=-fdt^{2}+\frac{dr^{2}}{f}+\beta rd\theta ^{2},  \label{5}
\end{equation}%
where $f=8\Lambda \beta r-2m\sqrt{r}+8Q^{2}$, which can be rewritten as

\begin{equation}
f=8\Lambda \beta \left( \sqrt{r}-r_{+}\right) \left( \sqrt{r}-r_{-}\right),
\label{6n}
\end{equation}

in which $r_{+}$ and $r_{-}$\ correspond to the event (outer) and inner
horizons of the charged MSW black holes:

\begin{equation}
r_{\pm }=\frac{m\pm \sqrt{m^{2}-64\Lambda \beta Q^{2}}}{8\Lambda \beta }.
\label{7n}
\end{equation}%
For the MSW black hole \cite{ChanMann}, the dilaton field is given $\phi=-\frac{1}{4}ln(\frac{r}{\beta})$ and the non-zero Maxwell tensor components are $F_{tr}=-F_{rt}=e^{4\phi}\frac{Q}{\sqrt{\beta{r}}}=\frac{Q\sqrt{\beta}}{r^{\frac{3}{2}}}$ in which $Q$ denotes the electric charge belonging to the electromagnetic vector potential $A_{\mu}=2Q\sqrt{\frac{\beta}{r}}\delta{_\mu^t}$. To have a black hole solution, ${m}\geqslant 8Q\sqrt{\Lambda \beta }$. For the uncharged MSW black holes (at the limit of $Q\rightarrow 0$), the horizons
reduce to $r_{+}=\frac{m}{4\Lambda \beta }$ and $r_{-}=0$.

Using Eq. (\ref{7n}), one can obtain the Hawking temperature of the MSW
black holes via \cite{msw5}
\begin{equation}
T_{H}=\left. \frac{1}{4\pi }\frac{d}{dr}\left( -g_{tt}\right) \sqrt{%
-g^{tt}g^{rr}}\right\vert _{\sqrt{r}=r_{+},}  \label{8n}
\end{equation}%
which results in

\begin{equation}
T_{H}=\frac{8\Lambda \beta r_{+}-m}{4\pi r_{+}}=\frac{\sqrt{m^{2}-64\Lambda
\beta Q^{2}}}{4\pi r_{+}}{.}  \label{9n}
\end{equation}

At the $Q=0$ limit, the Hawking temperature of an uncharged MSW black hole
becomes $\left. T_{H}\right\vert _{Q=0}=\frac{\Lambda \beta }{\pi }$, which
means that having an ordinary black hole (with positive-valued temperature)
is conditional on $\Lambda \beta \geq 0$. Moreover, it is obvious from Eq.
(8) that for a real physical temperature of the charged MSW black hole, the
condition of ${m}\geqslant 8Q\sqrt{\Lambda \beta }$ should also be
stipulated. On the other hand, it is worth noting that a negative
temperature is a well-known issue in spin systems with some upper energy
level limits \cite{Kittel}, and corresponds to a non-blackbody spectrum for 
exotic black holes \cite{Park}.

\section{Stability of MSW Black Holes}

Our purpose is to explore the linear stability of the charged MSW black
holes. To this end, we borrowed some ideas from \cite{Clem} by considering the breakneck perturbation: the $s$-mode.

The electrically charged circularly symmetric static solution of the
Einstein-Maxwell-dilaton field equations can be described by

\begin{equation}
ds^{2}=-\exp (2\gamma )dt^{2}+\exp (2\alpha )dr^{2}+\exp (2\eta )d\theta
^{2},  \label{10n}
\end{equation}

with the Maxwell tensor component

\begin{equation}
F_{tr}=q\exp \left( \alpha +\gamma -\eta +4\phi \right) ,  \label{11n}
\end{equation}

where $\gamma $, $\alpha $, and $\eta $ are ($r$,$t$) dependent metric
functions and $q=Q$. The field equation (\ref{2}) admits the following
non-zero $R_{tr}$ and $R_{\theta }^{\theta }$ components:%
\begin{equation}
R_{tr}=R_{rt}=4\overset{.}{\phi }\phi ^{^{\prime }},  \label{12n}
\end{equation}

\begin{equation}
R_{\theta }^{\theta }=-2\left[ \Lambda -q^{2}\exp (-2\eta )\right] \exp
(4\phi ),  \label{13n}
\end{equation}

where prime and dot denote derivatives with respect to $r$ and $t$,
respectively. One can assume that the metric functions and the dilaton field
have the following small perturbations:

\begin{equation}
\gamma \equiv \gamma \left( r,t\right) =\gamma _{0}\left( r\right) +\in
\gamma _{1}\left( r,t\right) ,  \label{14n}
\end{equation}

\begin{equation}
\alpha \equiv \alpha \left( r,t\right) =\alpha _{0}\left( r\right) +\in
\alpha _{1}\left( r,t\right) ,  \label{15n}
\end{equation}

\begin{equation}
\eta \equiv \eta \left( r,t\right) =\eta _{0}\left( r\right) +\in \eta
_{1}\left( r,t\right) ,  \label{16n}
\end{equation}

\begin{equation}
\phi =\phi \left( r,t\right) =\phi _{0}\left( r\right) +\in \phi _{1}\left(
r,t\right) .  \label{17n}
\end{equation}

Setting $\eta _{1}\left( r,t\right) =0$ gauge and comparing the metrics (\ref%
{5}) and (\ref{10n}), one obtains the following identity: $\exp (2\eta
)=\beta{r}$. Thus, after a straightforward calculation, $R_{tr}$ and $%
R_{\theta }^{\theta }$ components of the Ricci tensor and the Klein-Gordon
equation (\ref{4}) become

\begin{equation}
R_{tr}=\overset{.}{\alpha }\eta ^{^{\prime }},  \label{18n}
\end{equation}

\begin{equation}
R_{\theta }^{\theta }=\left[ \eta ^{^{\prime }}(\alpha ^{^{\prime }}-\gamma
^{^{\prime }})-(\eta ^{^{\prime }})^{2}-\eta ^{^{\prime \prime }}\right]
\exp \left( -2\alpha \right) ,  \label{19nn}
\end{equation}

\begin{equation*}
\exp (-2\alpha )\left[ \phi ^{^{\prime \prime }}-\phi ^{^{\prime }}(\alpha
^{^{\prime }}-\gamma ^{^{\prime }}-\eta ^{^{\prime }})\right] -\exp
(-2\gamma )\left[ \overset{..}{\phi }+\overset{.}{\phi }(\overset{.}{\alpha }%
-\overset{.}{\gamma })\right] +
\end{equation*}

\begin{equation}
\exp (4\phi )\left[ -\Lambda -q^{2}\exp (-2\eta )\right] =0.  \label{20n}
\end{equation}

Matching Eqs. (\ref{12n}) and (\ref{13n}) with Eqs. (\ref{18n}) and (\ref%
{19nn}) and performing the perturbations considering Eqs. (\ref{14n}-\ref%
{17n}) with the gauge of $\eta _{1}\left( r,t\right) =0$, the linearized
forms of the Einstein equations for the $R_{tr}$ and $R_{\theta }^{\theta }$
components and the Klein-Gordon equation (\ref{20n}) yield the following
expressions:
\begin{equation}
\overset{.}{\alpha _{1}}+2\overset{.}{\phi _{1}}=0,  \label{20old}
\end{equation}

\begin{equation}
4\left( \Lambda \beta r-Q^{2}\right) \left( \alpha _{1}+2\phi _{1}\right)
+r\left( 4\Lambda \beta r-m\sqrt{r}+4Q^{2}\right) (\alpha _{1}^{^{\prime
}}-\gamma _{1}^{^{\prime }})=0,  \label{21n}
\end{equation}

\begin{equation*}
(\alpha _{1}^{^{\prime }}-\gamma _{1}^{^{\prime }}+4r\phi _{1}^{^{\prime
\prime }})(4\Lambda \beta r-m\sqrt{r}+4Q^{2})+4(\alpha _{1}+2\phi
_{1})\left( \Lambda \beta -\frac{Q^{2}}{r}\right)-
\end{equation*}

\begin{equation}
\frac{r}{(4\Lambda \beta r-m\sqrt{r}+4Q^{2})}\overset{..}{\phi _{1}}%
+4(6\Lambda \beta r-m\sqrt{r}+2Q^{2})\phi _{1}^{^{\prime }}=0.  \label{22n}
\end{equation}

Multiplying Eq. (\ref{22n}) by $r$ leads to
\begin{equation*}
4\left( \Lambda \beta r-Q^{2}\right) \left( \alpha _{1}+2\phi _{1}\right)
+r\left( 4\Lambda \beta r-m\sqrt{r}+4Q^{2}\right) (\alpha _{1}^{^{\prime
}}-\gamma _{1}^{^{\prime }})+4r^{2}\phi _{1}^{^{\prime }}\left( 4\Lambda
\beta r-m\sqrt{r}+4Q^{2}\right)
\end{equation*}%
\begin{equation}
-\frac{r^{2}}{(4\Lambda \beta r-m\sqrt{r}+4Q^{2})}\overset{..}{\phi _{1}}%
+4r(6\Lambda \beta r-m\sqrt{r}+2Q^{2})\phi _{1}^{^{\prime }}=0,  \label{23n}
\end{equation}

and the first two terms vanish due to Eq. (\ref{21n}). This implies that the
linearized Klein-Gordon equation is independent of the solution \big($\alpha _{1}=-2\phi _{1}+f(r)$\big) of Eq. (\ref{20old}), since Eq. (\ref{23n}) comprises Eq. (\ref{21n}).
Ultimately, one can get

\begin{equation}
\phi _{1}^{^{\prime \prime }}+\left[ \frac{6\Lambda \beta r-m\sqrt{r}+2Q^{2}%
}{r(4\Lambda \beta r-m\sqrt{r}+4Q^{2})}\right] \phi _{1}^{^{\prime }}-\frac{%
\overset{..}{\phi _{1}}}{(8\Lambda \beta r-2m\sqrt{r}+8Q^{2})^{2}}=0.
\label{24n}
\end{equation}

Using the Fourier transformation with respect to the time coordinate, one
can introduce

\begin{equation}
\phi _{1}\left( r,t\right) =\phi _{1}\left( r\right) \exp \left( -ikt\right)
,  \label{25n}
\end{equation}

where $k$ (real) represents the frequency. Therefore, Eq. (\ref{24n})
reduces to the following effective Klein-Gordon equation:

\begin{equation}
\phi _{1}^{^{\prime \prime }}\left( r\right) +h\phi _{1}^{^{\prime }}\left(
r\right) -j\phi _{1}\left( r\right) =0,  \label{26n}
\end{equation}

where $h\ $and $j$ are the functions of $r$, which are provided as

\begin{equation}
h=\frac{12\Lambda \beta r-2m\sqrt{r}+4Q^{2}}{rf},  \label{27n}
\end{equation}

\begin{equation}
j=\frac{-k^{2}}{f^{2}}.  \label{28n}
\end{equation}

One can introduce the tortoise coordinate as follows.

\begin{equation}
u\equiv{u(r)=\int \frac{dr}{f}= \frac{1}{4\Lambda\beta\left( r_{+}-r_{-} \right)}\ln  \left( {\frac { \left( \sqrt {r}-r_{+} \right) ^{r_{+}}}{ \left( \sqrt {r}-r_{-} \right) ^{r_{-}}}} \right). \label{29n}}
\end{equation}

The range $r_{+} < \sqrt{r} < \infty$ corresponds to $−\infty < u < \infty$ (since $u\rightarrow-\infty$ as $\sqrt{r} \rightarrow r_{+}$). As one approaches to the MSW black hole, the radial coordinate changes more and more slowly with $u$, since $\frac{d{r}}{du}\rightarrow0$. Hence, we only look at the spacetime outside the horizon. The tortoise coordinate (\ref{29n}) indicates that once the MSW black hole is extremal ($r_{+}=r_{-}$), the analysis performed cannot be straightforwardly applied; as $u$ becomes undefined. After some manipulations, one can show that Eq. (\ref{26n}) recasts in

\begin{equation}
\frac{d^2 \phi_{1}(u)}{du^2}  +X\frac{d\phi_{1}(u)}{du}+k^{2}\phi_{1}(u) =0,  \label{30n}
\end{equation}

where

\begin{equation}
X=4\Lambda \beta -\frac{m}{\sqrt{r}}+\frac{4Q^{2}}{r}.  \label{31n}
\end{equation}

Setting

\begin{equation}
\phi _{1}\left( u\right) =\mathcal{R}(u)r^{-\frac{1}{4}},  \label{32n}
\end{equation}

and substituting Eq. (\ref{32n}) into Eq. (\ref{30n}), after a
straightforward calculation, one can obtain a Schr\"{o}dinger like
one-dimensional wave equation (see for example Chandrasekhar’s exhaustive book \cite{Chandram}):

\begin{equation}
-\frac{d^2 \mathcal{R}}{du^2}  +\left[ V_{eff}(r)-k^{2}\right]
\mathcal{R}\left( u\right) =0,  \label{33n}
\end{equation}

where $V_{eff}(r)$ is the effective potential, given by

\begin{equation}
V_{eff}\left( r\right) =f\frac{[f+4(\sqrt{r}m-8Q^{2})]}{16r^{2}}.
\label{34n}
\end{equation}

\bigskip
As discussed in the study of Dennhardt and Lechtenfeld \cite%
{Dennhardt}, in order for proving the stability of the black hole spacetime, it is
required to show the nonexistence of the $k^{2}<0$ solution, i.e.
non-existence of an exponentially growing solution \cite{Kimura}. In the black hole case (${m}\geqslant 8Q\sqrt{%
\Lambda \beta }$), one can see that $V_{eff}\left( r\right) $ is positive
definite for all $\sqrt{r}>r_{+}$. This is because when $\sqrt{r}>r_{+},$ $f>0$, one obtains

\begin{equation}
\left. \sqrt{r}m-8Q^{2}\right\vert _{\sqrt{r}>r_{+}}>0.  \label{35n}
\end{equation}

Non-negative $V_{eff}\left( r\right) $ implies $k^{2}\geqslant 0$ so that
there are no bounded solutions \cite{Dennhardt}. Namely,
the positiveness of the effective potential governing the perturbations
forbids any unstable modes. Thus, the MSW black holes
are linearly stable in the $s$-sector.

\section{Conclusion}

We applied linear stability analysis to MSW black holes. For this purpose,
we followed a semi-analytical method used in \cite{Kanti,Clem} which is mainly
based on the Fubini-Sturm theorem \cite{Fubini}. Then, we considered small
spacetime-dependent perturbations in both $R_{tr}$ and $R_{\theta }^{\theta
} $ components of the Einstein equations and in the Klein Gordon equation
(4). The key feature of the applied method was reduction of the linearized
equations to a single one-dimensional Schr\"{o}dinger type differential
equation. As a result, the effective potential of the MSW black holes was
derived. In the black hole case (${m}\geqslant 8Q\sqrt{\Lambda \beta }$),
the effective potential (\ref{34n}) was found to be positive through $\sqrt{r}>r_{+}$.
This means that there are no bounded solutions of Eq. (\ref{33n}) \cite%
{Dennhardt}. Thus, we concluded that the MSW black hole solutions are linearly stable for the
potentially most dangerous perturbation, which is the $s$-mode. Our findings
give support to \cite{SMSW}, in which it was shown that MSW black holes are
stable against \textit{small time-dependent perturbation}.

Our future plan is to extend our linear stability analysis to higher
dimensional ($d\geq 3$) rotating black holes in Einstein-Maxwell-dilaton
gravity \cite{Sheykhi}. Through this, we aim to see the effects of rotation
parameters and the number of dimensions on the stability of black holes.

\acknowledgments 
We are thankful to the Editor and anonymous Referees for
their constructive suggestions and comments.

\end{document}